\documentclass[aps,twocolumn,raggedbottom,showpacs,amssymb,groupedaddress]{revtex4}

\usepackage[english]{babel}
\usepackage{graphicx}
\usepackage{amssymb}
\usepackage{amsmath}
\usepackage{amscd}
\usepackage{eucal}
\usepackage{color}
\usepackage{bm}

\def\be{\begin{equation}}
\def\ee{\end{equation}}
\def\bea{\begin{eqnarray}}
\def\eea{\end{eqnarray}}
\def\bsub{\begin{subequations}}
\def\esub{\end{subequations}}

\begin{document}

\title{Tunable thermopower in a graphene-based topological insulator}

\author{Oleksii Shevtsov}
\affiliation{CEA-INAC/UJF Grenoble 1, SPSMS UMR-E 9001, Grenoble F-38054, France}
\author{Pierre Carmier}
\affiliation{CEA-INAC/UJF Grenoble 1, SPSMS UMR-E 9001, Grenoble F-38054, France}
\author{Christoph Groth}
\affiliation{CEA-INAC/UJF Grenoble 1, SPSMS UMR-E 9001, Grenoble F-38054, France}
\author{David Carpentier}
\affiliation{CNRS - Laboratoire de physique, Ecole Normale sup\'erieure de Lyon, France}
\author{Xavier Waintal}
\affiliation{CEA-INAC/UJF Grenoble 1, SPSMS UMR-E 9001, Grenoble F-38054, France}
\date{\today}

\begin{abstract}
Following  the recent proposal by Weeks et al., which suggested that indium (or thallium) adatoms deposited on the surface of graphene should turn the latter into a quantum spin Hall (QSH) insulator characterized by a sizeable gap, we perform a systematic study of the transport properties of this system as a function of the density of randomly distributed adatoms. While the samples are, by construction, very disordered, we find that they exhibit an extremely stable QSH phase with no signature of the spatial inhomogeneities of the adatom configuration. We find that a simple rescaling of the spin-orbit coupling parameter allows us to account for the behaviour of the inhomogeneous system using a homogeneous model. This robustness opens the route to a much easier experimental realization of this topological insulator. We additionally find this material to be a very promising candidate for thermopower generation with a target temperature tunable from $1$ to $80\,\mathrm{K}$ and an efficiency $ZT\approx 1$.
\end{abstract}

\maketitle

\section{Introduction}
The unique states which exist at the boundary of topological phases are of fundamental interest, but also of potential interest for applications due to their robustness. Ideally, these edge states determine entirely the transport properties of the sample. Moreover, the topological nature of the bulk phase implies a robustness of these edge states with respect to \emph{e.g.}\ sample geometry, impurities or other external perturbations. This topological robustness constitutes a fascinating motivation for designing electronic or spintronic devices based on these phases, as opposed to more fragile materials like \emph{e.g.}\ graphene \cite{Rotenberg:2011}. Of particular interest are the recently discovered topological insulators which are the focus of much attention: their edge states possess unique spin transport properties, due to their fixed spin helicity imposed by the inherent strong spin-orbit (SO) interaction. The surface states of the three-dimensional realization of these topological insulators resemble the low-energy Dirac fermions of graphene but here with a single species (as opposed to four for graphene), which implies the expected topological robustness \cite{Fu07prl,Moore07,Roy09}. Similarly, the edge states of the two-dimensional QSH topological phase are helical, consisting of a Kramers pair of states carrying two opposite spin currents \cite{Kane05,Kane05bis}.

Unfortunately, up to now probing the edge states of these new phases through transport experiments has proven a major challenge. While the surface states of candidate materials for the three-dimensional phase have been identified through ARPES and STM experiments \cite{Hsieh08,chen:2009,alpichshev:2010}, transport experiments have been hampered by a large contribution of spurious bulk states which obscures any clear signature of the surface states \cite{Checkelsky:2009}. On the other hand, the two-dimensional QSH phase has only been discovered experimentally in epitaxially grown HgTe/CdTe quantum wells following the initial theoretical proposal \cite{Bernevig06,Konig07}. In this material, an unambiguous signature of the helical edge states was observed through fractional four-probe conductance measurements \cite{roth:2009}. However, only one experimental group has mastered the fabrication of HgTe quantum wells so far.

In this article, we focus on a recent proposal to induce the two-dimensional QSH phase on a much more readily available material: graphene. While the intrinsic SO interaction of graphene is too weak to open an observable gap following the initial Kane and Mele proposal \cite{Kane05,Kane05bis}, it is possible to increase it by extrinsic means, as was demonstrated in a recent experiment \cite{Crommie10} where the generation of very large pseudo-magnetic fields by applying mechanical strain was observed. A more recent proposal suggests that the topological insulating behavior can be induced by suitable heavy adatoms deposited on the surface of the graphene sheet: uniform coverage with indium (or thallium, which is unfortunately poisonous) was predicted to induce a QSH phase \cite{Weeks11} with an insulating gap of the order of $100\,\mathrm{K}$ for a few percent coverage by indium. Given the technological expertise that exists for graphene, the possibility of a graphene-based two-dimensional topological insulator opens formidable perspectives. Since the breakthrough scotch-tape experiment carried by the Manchester group \cite{Novoselov04}, the field of graphene science has grown to impressive proportions, largely due to the numerous promises this revolutionary material holds for many applications ranging from (opto-)electronics to chemistry \cite{Geim09,Bonaccorso10}. Graphene samples of good quality are now routinely available around the world.

In the present work, we explore the condition and domain of existence of the QSH phase in the realistic situation where adatoms are deposited at random on the surface of graphene. The signature of the QSH phase is found for an extremely wide range of parameters, signaling its robustness to inhomogeneous SO interaction. More importantly, we show that the transport properties of this phase show no sign of this inhomogeneity, and can be entirely described by an effective homogeneous theory which is valid for any adatom configuration (i.e., before \emph{and} after averaging over adatom configurations). We find that the resulting QSH phase shows remarkable tunable thermopower characteristics, in a range of temperatures where usual (narrow-band semi-conductor) materials lead to rather poor yields.

\section{Model}
We consider a tight-binding model describing the low-energy spectrum of graphene in the presence of adatoms \cite{Weeks11}:
\be
\label{eq:KM}
H = -t\sum_{\langle i,j \rangle,\alpha} c_{i,\alpha}^\dagger c_{j,\alpha}
+ i\lambda_{\text{so}}\sum_{\mathcal{P}}\sum_{ \langle\langle i,j\rangle\rangle \in \mathcal{P},\alpha,\beta}\nu_{ij} c_{i,\alpha}^\dagger s_{\alpha\beta}^z c_{j,\beta} \; .
\ee
Indices $i,j$ label the lattice sites, $\alpha,\beta$ the spin quantum numbers, $\langle\rangle$ stands for nearest neighbors, $\langle\langle\rangle\rangle$ for second nearest neighbors. The nearest-neighbor hopping $t\approx 3\,\mathrm{eV}$ will set the energy scale: hereafter all energies are expressed in units of $t$ while sizes are expressed in units of $\sqrt{3} a$ ($a$ is the carbon-carbon distance). The second term describes the dominant SO coupling induced by an indium (or thallium) adatom, with $s^z$ the usual Pauli matrix in spin space. The adatoms are distributed at random on a fraction $n_{\text{ad}}$ of the graphene hexagonal plaquettes $\mathcal{P}$ (see the lower inset of Fig.\ \ref{fig4T} for a cartoon of a plaquette). SO coupling induces chiral second nearest neighbor hoppings with $\nu_{ij}=1$ when moving counter-clockwise around a plaquette, and $-1$ otherwise. The intensity of the induced SO interaction is parametrized by $\lambda_{\text{so}}$ with $\lambda_{\text{so}}\approx 0.0067$ for indium adatoms (and $\lambda_{\text{so}}\approx 0.02$ for thallium).

The original Kane-Mele Hamiltonian \cite{Kane05,Kane05bis} for pure graphene is recovered by considering SO coupling on all plaquettes ($n_{\text{ad}}=1$). In this case, the uniform SO interaction opens a gap $2 \Delta_{\text{so}}=6\sqrt{3}\lambda_{\text{so}}$ in the vicinity of the Dirac points $K$, $K'$ of the Brillouin zone. The corresponding QSH phase possesses a $\mathbb{Z}_{2}$ topological symmetry, which manifests itself through the appearance of a pair of edge states at the boundary of a sample. The signature of this phase was found through ab-initio calculations in Ref.\ \cite{Weeks11}.

\section{Parametric study of the QSH phase}
\begin{figure}[]
\begin{center}
\includegraphics[angle=0,width=1.0\linewidth]{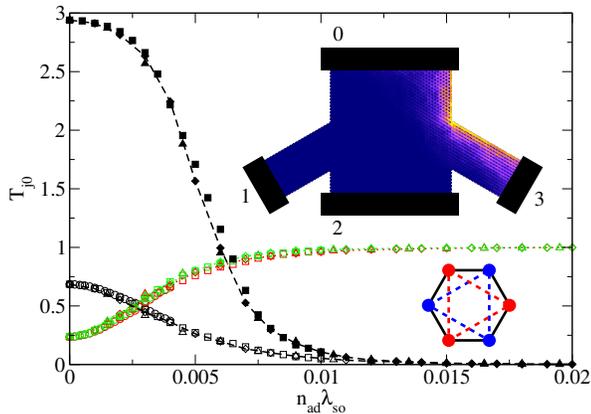}
\caption{(Color online): Upper inset: schematic of our setup: a $4$-terminal graphene cross with armchair edges and width of the small arm $W=22$. The different colors correspond to an actual calculation of the current density for spin-up electrons upon injection from contact $0$. The existence of an edge state is manifest. Main figure: Scaling of the transmissions $T_{j0}$ from contact $0$ for various couplings $\lambda_{\text{so}}$ and adatom densities $n_{\text{ad}}$, plotted as a function of the effective SO coupling strength $\lambda_{\text{so}}^{{\rm eff}}= \lambda_{\text{so}} n_{\text{ad}}$. Dashed (black) lines correspond to ``longitudinal" transmission $T_{20}$ and dotted lines correspond to ``Hall" transmissions $T_{10}$ (red) and $T_{30}$ (green). The two sets of curves correspond to an energy $E = 0$ (open symbols), and $E = 0.05$ (filled symbols). Different symbol shapes correspond to different values of SO coupling, $\lambda_{\text{so}} = 0.02$, $0.05$, $0.1$ and $0.15$ (because they all collapse on the same curve, various symbols may seem indistinguishable from each other). Note that no averaging over adatom configurations has been performed here. Lower inset: cartoon of a graphene plaquette. Full lines stand for direct hopping elements while dashed lines correspond to the SO-induced hopping elements.}
\label{fig4T}
\end{center}
\end{figure}

In order to monitor the existence of the QSH phase, we focus on the signature of the associated edge modes on the conductance matrix of a multi-terminal sample. The latter is given by the multi-terminal Landauer-B\"uttiker formula, which expresses the current $I_\alpha$ flowing in the electrode $\alpha$ (see upper inset of Fig.~\ref{fig4T} for a sketch of the sample) in presence of a potential $V_\beta$ as 
\be
I_\alpha=\frac{e^2}{h} \sum_\beta T_{\alpha\beta} V_\beta \; , 
\ee
where $T_{\alpha\beta}$ is the transmission coefficient between two electrodes. It can be computed numerically from the knowledge of the retarded Green function $G$ via the formula
\be
T_{\alpha\beta} = {\text{Tr}}[\Gamma_\alpha G\Gamma_\beta G^\dagger] \; ,
\ee
with $\Gamma_\alpha={\text{Im}}(\Sigma_\alpha)$, where $\Sigma_\alpha$ is the self-energy of lead $\alpha$, and with the Green function given by the expression
\be
G(E)=(E-H-\sum_\alpha \Sigma_\alpha)^{-1}
\ee
(with the Hamiltonian $H$ in its first quantization form). In the QSH phase the transmission between successive probes reads exactly $T_{\alpha,\alpha+1} = T_{\alpha,\alpha-1} = 1$, while all other transmission coefficients $T_{\alpha,\beta}$ ($\alpha\ne\beta$) vanish. In the following, we will use this unique characterization of the QSH phase in the 4-terminal cross geometry of Fig.~\ref{fig4T}. The 4-terminal geometry always provides an unambiguous characterization of this phase, as opposed to the two-terminal conductance where $g_{2T} = 2 e^{2}/h$ can be observed at the Dirac point independently of the presence of SO interaction.

We also pay special attention to avoid the appearance of spurious effects in nano-ribbon geometries. Structural boundary conditions of graphene are of two types: armchair and zigzag. The latter features a zero-energy edge state, signatures of which have been shown to obscure the appearance of the QSH edge state \cite{Metalidis11}. Therefore we will henceforth consider armchair-terminated graphene systems in order to avoid confusion between different types of edge states. Armchair ribbons are themselves divided into two families \cite{Brey06}: they can be either metallic when their width $W$ equals $2$ (modulo 3), or semi-conducting for other values of the width. In the following, we focus on metallic armchair ribbons to avoid the competition between the SO-interaction-induced gap and the finite-width-induced gap. We use a cross-like geometry, with a fixed aspect ratio (see inset of Fig.~\ref{fig4T}) and smallest width $W$ at contacts $1$ and $3$: the shape of the sample has been chosen such that most of the current is directly transmitted from $0$ to $2$ in the absence of SO interaction. The numerical calculations were performed using the KNIT package which implements a generalization of the recursive Green function algorithm to multi-terminal systems \cite{Kazymyrenko08}.

\begin{figure}[]
\begin{center}
\includegraphics[angle=0,width=1.0\linewidth]{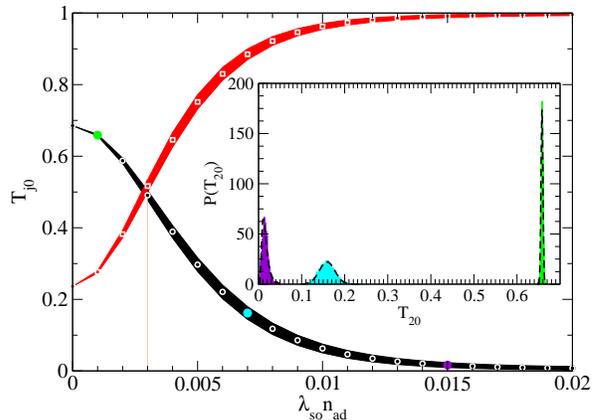}
\caption{(Color online): Statistics of the transmission coefficients $T_{20}$ (lower curve, black) and $T_{30}$ (upper curve, red) when averaged over 2500 adatom configurations, as a function of the effective SO coupling strength. The shaded regions represent one standard deviation from the mean value which is indicated by white symbols (circles for $T_{20}$ and squares for $T_{30}$). The corresponding density of probability for 3 points (green, cyan and violet filled circles, from left to right) are given in the inset: as the SO coupling increases, the probability distribution $P(T_{20})$ shifts from a Gaussian (right green and middle cyan) to a log-normal (left violet). Fits are indicated by dashed lines. All data points were generated for an energy $E=0$, a fixed adatom density $n_{\text{ad}}=0.1$, and the same system sizes as in Fig.~\ref{fig4T}.}
\label{figStat}
\end{center}
\end{figure}

Fig.~\ref{fig4T} shows the ``longitudinal" $T_{20}$ and ``Hall" $T_{10}$, $T_{30}$ transmission coefficients as a function of the strength of the SO interaction and for various concentrations $n_{\text{ad}}$ of adatoms. One observes, as expected, that upon increasing the strength of $\lambda_{\text{so}}$ for a fixed concentration $n_{\text{ad}}$ of adatoms, one enters the QSH phase: the Hall coefficients tend to unity while the longitudinal one vanishes. The upper inset shows the actual up spin current density inside the sample in the QSH regime: we recover the expected edge state characteristics of the QSH phase, including an exponential decay of the current as a function of the distance to the edge but also oscillations with frequency $|{\bf K} - {\bf K'}|$ coming from the valley-mixing armchair boundary condition \cite{Metalidis11}. The crucial point shown in Fig.~\ref{fig4T} is that \emph{all the results} are rescaled as a function of the effective SO interaction strength
\be
\lambda_{\text{so}}^{{\rm eff}} = \lambda_{\text{so}} n_{\text{ad}} \; .
\ee
In other words, the QSH phase, while originating from a very inhomogeneous sample, is perfectly described in \emph{each sample} by an effective homogeneous phase with a uniform coverage but a reduced SO strength. In this phase, each adatom occupying one of the plaquettes spreads its coupling $\lambda_{\text{so}}$ over a distance $\xi_{\text{ad}}=1/\sqrt{n_{\text{ad}}}$, resulting in a weaker effective SO interaction strength $\lambda_{\text{so}}^{{\rm eff}} = \lambda_{\text{so}} /\xi_{\text{ad}}^{2}$ but a uniform effective full coverage $n_{\text{ad}}=1$. The surprising occurrence of this mean-field description in each disordered sample and down to very small concentrations can be attributed to the large localization length $\xi$ of the QSH edge states (see below). This large value of $\xi$ also explains why rather large values of $\lambda_{\text{so}}^{{\rm eff}}$ are required to observe perfect ``Hall" transmissions in the small samples considered in Figs.~\ref{fig4T} and \ref{figStat}. In real, $\mu m$-sized samples, the required value of $\lambda_{\text{so}}^{{\rm eff}}$ will be much smaller since deviations to perfect "Hall" effect are controlled by the ratio $W/\xi$ between the width of the small arm $W$ and the localization length $\xi$ (see Fig.~\ref{figXi} below).

We have found that the above \emph{effective homogeneous description} is very robust and applies for (i) a single realization of the adatom configuration, i.e.\ does not require disorder averaging, (ii) various sizes from wide samples deep into the QSH phase down to narrow samples where the edge states on both sides of the sample have a finite overlap, (iii) different energies, from the Dirac point up to the SO-interaction-induced gap. To illustrate this robustness, we performed some statistics and show in Fig.~\ref{figStat} how the transmission coefficients vary from one adatom configuration to another. The ``envelope" curves displayed in Fig.~\ref{figStat} correspond to the average transmission plus or minus one standard deviation (i.e. for a Gaussian distribution there is a 68\% probability that the outcome will fall inside the envelope for a given adatom configuration). At small SO coupling, the system is essentially ballistic with very small (Gaussian) fluctuations. At large SO coupling, the system is deep in the QSH phase with small (log-normal) fluctuations (the inset of Fig.~\ref{figStat} shows the actual probability distributions). We find that, in the crossover between these two limits, the fluctuations remain remarkably low for all values of the effective SO coupling strength. Additionally, we saw no evidence for a breakdown of our mean-field description, even down to very low adatom concentrations.

A celebrated property of a topological phase is its robustness with respect to the presence of disorder.
\begin{figure}[]
\begin{center}
\includegraphics[angle=0,width=1.0\linewidth]{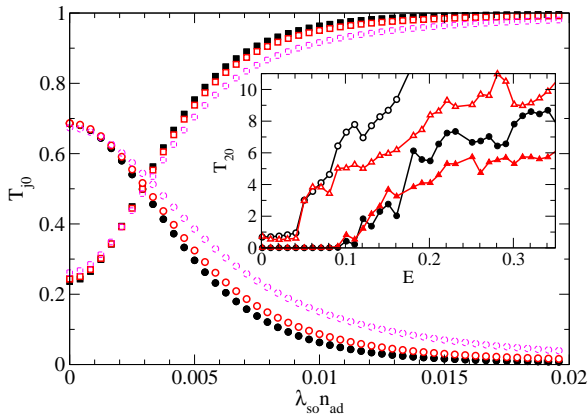}
\caption{(Color online): Transmission coefficients $T_{20}$ (circles) and $T_{30}$ (squares) as a function of the effective SO coupling strength, for various values of onsite disorder: $V=0$ (filled black), $V=0.4$ (empty red) and $V=0.8$ (empty dashed magenta). No qualitative change is brought about by disorder, although some deviation from the $V=0$ curve in the crossover region between the metal and the QSH phases starts to be visible at strong enough ($V=0.8$) disorder. The data were averaged over 50 distinct realizations of disorder and adatom configuration for each value of $\lambda_{\text{so}}^{{\rm eff}}$. The energy was kept at $E=0$, and the system sizes as in Fig.~\ref{fig4T}. Inset: Longitudinal transmission $T_{20}$ as a function of energy for $\lambda_{\text{so}}^{{\rm eff}}=0$ (empty symbols) and $\lambda_{\text{so}}^{{\rm eff}}=0.02$ (filled symbols). Circles are for $V=0$ and triangles for a single disorder configuration with $V=0.8$. Disorder has no effect on $T_{20}$ in the QSH phase ($E<\Delta_{\text{so}}\approx0.1$).}
\label{figDis}
\end{center}
\end{figure}
The persistence of this property in the present context can easily be checked by adding on-site disorder (on each site) to our Hamiltonian in Eq.~(\ref{eq:KM}), following the standard prescription 
\be
H_{\text{dis}} = \sum_{i,\alpha} V_ic_{i,\alpha}^\dagger c_{i,\alpha} \; ,
\ee
where $V_i$ is a disorder strength randomly distributed in the interval $[-V/2,V/2]$. The transmission coefficients as a function of the effective spin-orbit coupling for several values of $V$ are shown in Fig.~\ref{figDis}. As expected for a topological phase, the presence of disorder provides no qualitative (and hardly any quantitative) modification to the general picture described earlier. Deviations from the ``clean" ($V=0$) case remain small, unless the strength of disorder reaches (extremely strong) values of the order of the hopping parameter $t$. Note that, in the main panel of Fig.~\ref{figDis}, the Fermi energy is very close to the Dirac point. It is well established that disorder has a very small effect on the metallic phase (at small $\lambda_{\text{so}}$) close to the Dirac point \cite{mirlin09,richter11}. This point can be understood from a semi-classical consideration: close to the Dirac point, the Fermi wave length diverges so that only very long range disorder affects the physics.

Away from the Dirac point, onsite disorder does affect the transport properties of graphene, but not of the QSH phase. This is best seen in the inset of Fig.~\ref{figDis} where we plot the longitudinal transmission $T_{20}$ as a function of the Fermi energy in the presence/absence of spin-orbit coupling and disorder (for a typical sample). When spin-orbit coupling is present, adding disorder has no effect on $T_{20}$ so long as $E<\Delta_{\text{so}}\approx 0.1$, i.e. inside the QSH phase. In the absence of spin-orbit coupling (or above the spin-orbit gap), however, disorder strongly affects $T_{20}$ except at small energies close to the Dirac point. In short, we find that the QSH phase is very resilient to
onsite disorder.

\begin{figure}
\begin{center}
\includegraphics[width=1.0\linewidth]{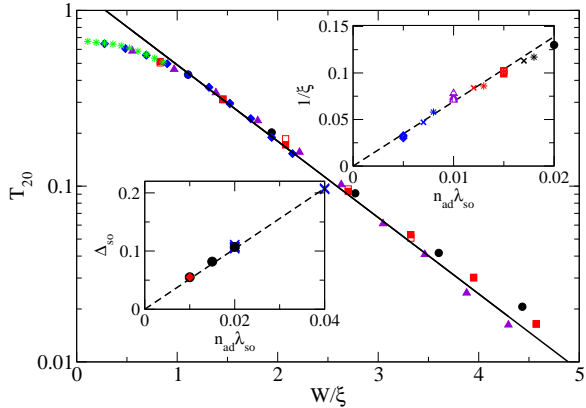}
\caption{(Color online): Averaged longitudinal transmission as a function of $W/\xi$ for energy $E = 0$. Filled symbols: fixed $\lambda_{\text{so}} = 0.1$ with $n_{\text{ad}}=0.2$ (circles), $0.15$ (squares), $0.1$ (triangles), $0.05$ (diamonds). Open symbols: idem but the role of $\lambda_{\text{so}}$ and fixed $n_{\text{ad}}$ are exchanged, stars: $\lambda_{\text{so}} = 0.02$. Line: $Y\propto\exp -W/\xi$ where $\xi$ is given by Eq.~\eqref{eq:xi}. Upper inset: $1/\xi$ (extracted from the data of the main plot) as a function of $\lambda_{\text{so}}^{{\rm eff}}$ including additional points for different energy $E = 0.02$,
different aspect ratio of the sample,
and more values of $n_{\text{ad}}$ 
and $\lambda_{\text{so}}$ (various symbols).
Dashed line: Eq.~\eqref{eq:xi}. Lower inset: $\Delta_{\text{so}}$ as a function of $\lambda_{\text{so}}^{{\rm eff}}$ for $\lambda_{\text{so}}=0.02$, $0.05$ and $0.1$ (various symbols). Dashed line: Eq.~\eqref{eq:gap}. In all cases, error bars due to sample-to-sample fluctuations are smaller than the symbol sizes.}
\label{figXi}
\end{center}
\end{figure}

The above considerations lead us to predict that for the present system, the physics of the QSH phase is entirely described by the Kane-Mele model (i.e.\ the full coverage case, $n_{\text{ad}}=1$) provided one performs the substitution $\lambda_{\text{so}}\rightarrow \lambda_{\text{so}} n_{\text{ad}}$. In particular, the known analytical expressions for the characteristic scales of the Kane-Mele model should apply here. We now explicitly check this for the gap $\Delta_{\text{so}}$ \cite{Kane05bis} of the QSH phase and the width $\xi$ \cite{Metalidis11} of the edge states, whose expected expressions read
\be
\label{eq:gap}
\Delta_{\text{so}} = 3\sqrt{3}\lambda_{\text{so}}^{{\rm eff}} \; ,
\ee
\be
\label{eq:xi}
\xi = \frac{\hbar v_F}{2\Delta_{\text{so}}} = \frac{1}{12\lambda_{\text{so}}^{{\rm eff}}} \; .
\ee
Fig.~\ref{figXi} shows the behavior of the longitudinal transmission $T_{20}$ (averaged over several adatom configurations) as a function of the width $W$. Its exponential decrease $T_{20}\propto e^{-W/\xi}$ allows for an accurate determination of $\xi$: as the width increases, the overlap between two opposite edge modes and the associated backscattering decrease. An extremely good scaling of the data is obtained with an expression of the transmission deduced from Eq.~\eqref{eq:xi}, confirming the above effective homogeneous description.

As a last test, we now turn to a study of the two-terminal transmission $T(E)$ of a rectangular graphene ribbon as a function of the Fermi energy $E_F$.
\begin{figure}
\begin{center}
\includegraphics[width=1.0\linewidth]{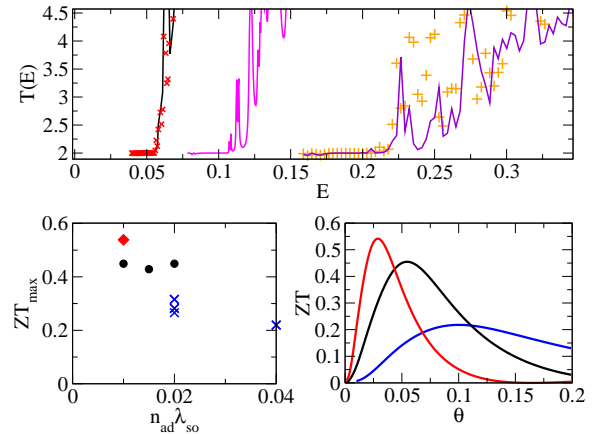}
\caption{(Color online): Upper plot: Transmission $T$ as a function of the energy $E$ in a simple armchair ribbon of width W, for $\lambda_{\text{so}}^{{\rm eff}}=0.01$ (left black line: $n_{\text{ad}}=0.2$, crosses: $n_{\text{ad}}=0.5$) with $W=130$, $\lambda_{\text{so}}^{{\rm eff}}=0.02$ (middle magenta line) with $W=65.5$, and $\lambda_{\text{so}}^{{\rm eff}}=0.04$ (right violet line and pluses: two different samples) with $W=32.5$. Lower right: $ZT$ as a function of temperature $\theta$ for $\lambda_{\text{so}}^{{\rm eff}}=0.01$, $0.02$ and $0.04$ from right to left. The Fermi energy was chosen to be equal to $1.5 \Delta_{\text{so}}$. Lower left: peak value $ZT_{\rm max}$ of $ZT(\theta)$ as a function of $\lambda_{\text{so}}^{{\rm eff}}$ for $\lambda_{\text{so}}=0.02$ (diamond), $\lambda_{\text{so}}=0.05$ (circles) and $\lambda_{\text{so}}=0.1$ (crosses).}
\label{figZT}
\end{center}
\end{figure}
The results are shown in Fig.~\ref{figZT} (upper plot): for energies inside the topological gap ($E_F\le\Delta_{\text{so}}$), the current is entirely carried by the edge states leading to $T=2$. Above the gap ($E_F>\Delta_{\text{so}}$), one leaves the QSH phase and the transmission increases quickly, allowing for a precise extraction of the gap value. Note that as $\Delta_{\text{so}}$ gets smaller, wider samples must be used to keep the number of open channels at $E_F=\Delta_{\text{so}}$ larger than one. In the lower inset of Fig.~\ref{figXi}, we plot the extracted $\Delta_{\text{so}}$ as a function of $\lambda_{\text{so}}^{{\rm eff}}$ for various values of $n_{\text{ad}}$ and find, again, a remarkable agreement with the effective homogeneous Kane-Mele description.

\section{Thermopower}
We now turn to an analysis of a very peculiar feature of Fig.~\ref{figZT}: the slope of the two-terminal transmission $T(E)$ above the gap is very steep, \emph{and gets steeper} as the gap decreases. This property has strong implications in terms of thermopower generation. To discuss thermopower at finite temperature within the Landauer-B\"uttiker framework, let us start by introducing
\be
T_n=\int dE \left(-\frac{\partial f}{\partial E}\right) (E-E_F)^n \; T(E),
\ee
where $f(E)=1/(e^{(E-E_F)/(k_B\theta)} + 1)$ is the Fermi function at temperature $\theta$, from which we can express the conductance $g=(e^2/h) T_0$, the heat conductance $g_H=1/(h\theta) T_2$ and the Seebeck conductance $g_S=-e/(h\theta) T_1$. The Seebeck coefficient $S=-\delta V/\delta\theta$ which characterizes the voltage $\delta V$ across a sample produced by a difference of temperature $\delta\theta$
between sample edges is simply given by $S=g_S/g$. The dimensionless figure of merit which measures the efficiency of the thermopower generator is known as the $ZT=\theta S^2 \sigma/\kappa$ parameter (we keep the standard notation $ZT$ although we call the temperature $\theta$ and not $T$). $\sigma$ ($\kappa$) is the electrical (thermal) conductivity. At low $ZT\ll 1$, a thermocouple has an efficiency equal to the fraction $ZT/4$ of the Carnot efficiency $\delta\theta/\theta$, while it tends to the Carnot value at large $ZT\gg 1$. Within the Landauer-B\"uttiker approach, we have \cite{sivan1986}
\be
\label{ZT1}
ZT=\frac{(T_1)^2}{T_0 T_2-(T_1)^2 }
\ee
which simplifies at low temperature into the so-called Mott formula \cite{cutler1969}
\be
\label{ZT2}
ZT=\frac{\pi^2}{3} (k_B\theta)^2 \left(\frac{T'(E_F)}{T(E_F)}\right)^2.
\ee
Hence the key towards an efficient thermocouple lies in low values of the transmission and simultaneously a steep variation of this transmission with the energy. The shape of the transmission versus energy curve $T(E)$ presented in Fig.~\ref{figZT} possesses all the required properties: the low transmission value $T=2$ accounts for the single edge state conducting channel (moreover it does not scale with the sample width $W$) while its derivative is typically of order $T'\approx 1/\Delta_{\text{so}}$. This results in very high values of $ZT\approx 1$ for temperatures $\theta\approx\Delta_{\text{so}}/k_B$.

To quantitatively describe the expected behavior, we have plotted the parameter $ZT$ as a function of the temperature $\theta$ in the lower right of Fig.~\ref{figZT}. We find well-defined maxima for $\theta\approx\Delta_{\text{so}}/(2k_B)$, while the optimal efficiency is reached for energy values $E_F \approx 3\Delta_{\text{so}}/2$. Note that the present material possesses the unique characteristics that the optimum temperature $\Delta_{\text{so}}/(2k_B)$ can be simply tuned by changing the concentration of adatoms while the Fermi level can be easily switched to the optimum value by \emph{e.g.}\ a simple back gate. The lower left plot of Fig.~\ref{figZT} shows the value of the peak of $ZT(\theta)$ as a function of $\lambda_{\text{so}}^{{\rm eff}}$: we find very high values, up to $ZT=0.5$, which \emph{tend to increase} upon decreasing the gap $\Delta_{\text{so}}$. Hence, these graphene-based topological insulators appear as very good candidates for low-temperature thermocouples: for instance a $6\%$ indium coverage is expected to give a gap of $80\,\mathrm{K}$, hence an optimum working temperature of $40\,\mathrm{K}$. At $1\%$ coverage, the optimum temperature lies around $10\,\mathrm{K}$, a target temperature for \emph{e.g.}\ radioisotope thermoelectric generators of spacecrafts for which a material with such high $ZT\approx 0.5$ would constitute a great improvement.

Finally, we note that the above estimate of $ZT$ only takes into account the electron contribution to the thermal conductivity. Real values should renormalize by a factor $\kappa/(\kappa+\kappa_{ph})$ where $\kappa_{ph}$ is the phonon contribution and should therefore be slightly smaller than predicted above. However, adding structural or extrinsic disorder to the ribbon or intentionally damaging the ribbon by making holes in it can drastically reduce the phonon-mediated thermal conductivity, while the \emph{topologically protected} physics discussed in this paper should remain largely unaffected.

\section{Conclusion}
One of the most important aspects associated with the discovery of graphene is the relative easiness (and low cost) of sample production. While very few groups can produce high mobility two-dimensional electron gases in semiconductor heterostructures, graphene physics requires lighter equipment and is being studied by an increasing number of groups. The same remark is even more applicable to the study of two-dimensional topological insulators. Hence, the proposal \cite{Weeks11} discussed in this paper for the realistic situation of inhomogeneous samples appears all the more promising.

We have shown that such a QSH phase, while originating from very inhomogeneous samples, can be described in \emph{each sample} by the known results on the pure Kane-Mele model with an effective SO coupling strength accounting for the density of adatoms. This description should hold in the presence of disorder or charge-density fluctuations induced by the underlying substrate, as long as the magnitude of these effects remains smaller than the gap $\Delta_{\text{so}}$. We have further demonstrated that this new material and its associated QSH phase provide a very efficient thermocouple at low temperature. Indeed, while important efforts have been made to improve the $ZT$ parameter at room temperature (values of $ZT>1$ can now be found), the existing materials are very inefficient at low temperature. For instance, existing thermocouples working around $10\,\mathrm{K}$ have $ZT<0.01$. Moreover, the graphene-based thermocouples discussed here have a target temperature which can be tuned from a few hundred degrees Kelvin ($n_{\text{ad}}=1$) down to $0\,\mathrm{K}$ ($n_{\text{ad}}=0$) by simply changing the concentration of adatoms while retaining extremely high values of $ZT\approx 1$.

\begin{acknowledgments}
This work was supported by EC Contract ERC MesoQMC, STREP ConceptGraphene as well as ANR 2010-Blanc IsoTop.
\end{acknowledgments}

\bibliography{thermoTI}
\bibliographystyle{apsrev}

\end{document}